\documentclass[usenatbib,useAMS,usegraphicx]{mn2e}
\usepackage{amsmath}
\usepackage{rotating}
\usepackage{times}
\usepackage{color}
\usepackage{subfigure}
\newcommand{\lens}{B1938+666}

\title[SHARP I - JVAS B1938+666]{SHARP - I. A high-resolution multi-band view of the infra-red Einstein ring of JVAS B1938+666}

\author[D. J. Lagattuta et al.]{D. J. Lagattuta,$^{1,2,3}$\thanks{dlagattu@astro.swin.edu.au} S.
  Vegetti,$^4$ C. D. Fassnacht,$^1$ M. W. Auger,$^{5,6}$
  L. V. E. Koopmans$^7$ \newauthor and J. P. McKean$^8$\\ 
  $^1$Department of Physics, University of California, Davis, 1
    Shields Avenue, Davis, CA 95616\\
  $^2$Centre for Astrophysics \& Supercomputing, Swinburne University of Technology,
      Hawthorn, VIC 3122, Australia\\
  $^3$ARC Centre of Excellence for All-sky Astrophysics (CAASTRO)\\
  $^4$Kavli Institute for Astrophysics and Space Research, Massachusetts
    Institute of Technology, Cambridge, MA 02139\\
  $^5$Department of Physics, University of California, Santa Barbara,
      CA 93106\\
  $^6$Institute of Astronomy, University of Cambridge, Madingley Road, 
      Cambridge CB3 0HA\\
  $^7$Kapteyn Astronomical Institute, University of Groningen, P.O. Box
      800, 9700 AV Groningen, The Netherlands\\
  $^8$Netherlands Institue for Radio Astronomy (ASTRON), Oude Hoogeveensedijk 4, 7991 PD Dwingeloo, The
      Netherlands\\}

\begin{document}

\date{Accepted 2012 May 28. Received 2012 May 24; in original form 2012 May 03}

\pagerange{\pageref{firstpage}--\pageref{lastpage}} \pubyear{2012}

\maketitle

\begin{abstract}
We present new mass models for the gravitational lens system \lens,
using multi-wavelength data acquired from Keck adaptive optics (AO)
and \emph{Hubble Space Telescope} (\emph{HST}) observations.  These
models are the first results from the Strong-lensing at High Angular
Resolution Program (SHARP), a project designed to study known
quadruple-image and Einstein ring lenses using high-resolution
imaging, in order to probe their mass distributions in unprecedented
detail.  Here, we specifically highlight differences between AO- and
\emph{HST}-derived lens models, finding that -- at least when the lens
and source galaxies are both bright and red, and the system has a high
degree of circular symmetry -- AO-derived models place significantly
tighter constraints on model parameters.  Using this improved
precision, we infer important physical properties about the
\lens\ system, including the mass density slope of the lensing galaxy
($\gamma = 2.045$), the projected dark matter mass fraction within the
Einstein radius ($M_{\rm dark}/M_{\rm lens} = 0.55$), and the total
magnification factor of the source galaxy ($\sim 13$).  Additionally,
we measure an upper-limit constraint on luminous substructure ($\rm
M_V > −16.2$), based on the non-detection of bright satellite galaxies
in all data sets. Finally, we utilize the improved image resolution of
the AO data to reveal the presence of faint arcs outside of the
primary Einstein ring.  The positions and orientations of these arcs
raise the intriguing possibility that \lens\ has a second source
galaxy, located at a more distant redshift.  However, future work is
needed to verify this hypothesis.
\end{abstract}

\begin{keywords}
galaxies: individual (JVAS B1938+666) --- gravitational lensing:
strong
\end{keywords}

\section{Introduction}

An understanding of the nature and distribution of matter on small
($<$ 1 Mpc) scales is essential to modern astrophysics.  Measuring the
shapes of galaxy mass profiles reveals the presence of dark
matter
\citep[e.g.,][]{rub80,van86,deb97,bos99,gav07,con07,dut11,ruff11,suy12}
and its interactions with baryonic matter
\citep[e.g.,][]{blu86,gne04,aug10,sch10}.  Separating total mass into
luminous and dark components can constrain cosmological parameters
(e.g. the baryon fraction $\Omega_b$) and place estimates on the
efficiency of star-formation in galaxies
\citep[e.g.,][]{fuk98,hey06,nap10,lag10a}.  Observing changes in any
of these quantities over cosmological time is a key component in
studying galaxy evolution \citep[e.g.,][]{man06,beh10,lag10a}.

While there are many techniques capable of measuring mass on small
scales, gravitational lensing stands out as an especially powerful
choice.  Unlike other methods, lensing directly measures a total
(baryonic + dark matter) mass without requiring this mass to be
luminous or in any specific dynamical state.  Furthermore, a lensing
analysis is not limited to the local Universe, but rather can be
applied to systems located over a wide range of cosmological
distances.  Strong gravitational lensing, in particular, can provide a
wealth of information about the nature of galaxies (e.g.,
\citealt{koc06} and references therein).  With typical image
separations of $\sim$ 1~arcsec, galaxy-scale strong lenses produce
mass estimates close to the centre of galaxies, giving information
about both the baryon-dominated luminous core and the inner regions of
the dark matter halo.  This relatively small angular size
(corresponding to physical scales of 5-10 kpc at typical
lensing-galaxy redshifts) decreases the probability of foreground
interlopers contaminating the line of sight {--} a problem that can
bias mass estimates obtained from group and cluster-scale strong
lenses, and can strongly dilute the signal measured from weak lensing
{--} suggesting that the observed lensing signal will be dominated by
the mass of the lensing galaxy.

In this paper, we investigate the galaxy-scale gravitational lens
\lens.  First discovered as part of the Jodrell Bank--Very Large Array
Astrometric Survey (JVAS; \citealt{pat92,bro98,wil98}), initial radio
observations of \lens\ showed a quadruply-imaged background source,
configured into a partial Einstein ring, along with a second,
doubly-imaged component \citep{pat92,kin97}.  Follow-up imaging in the
near-infrared (NIR) and optical uncovered a bright red object that was
thought to be the galaxy lensing the radio emission
\citep{rho96}. This emission was later shown by NIR \textit{Hubble
  Space Telescope} (\textit{HST}) imaging to be composed of the
light from both the lensing galaxy and a complete Einstein ring of the
background source galaxy \citep{kin98}.  \citet{ton00} measured a
redshift of $z_l =$~0.881 for the lensing galaxy from optical
spectroscopy, while \citet{rie11} determined a redshift of $z_s
=$~2.059 for the source from CO observations.

As \lens\ has an Einstein ring, its lensing mass model can be
determined to high precision.  \citet{koc01} showed that, when
compared with two-image quasar or arc lens systems, constraints from
an Einstein ring can be used to break degeneracies between the
monopole moment of the gravitational potential and higher order terms.
Removing these degeneracies allows for a more robust measurement of
the slope of the lensing galaxy's mass profile and, when combined with
time delay information, can provide an unambiguous measurement of the
Hubble constant (e.g. \citealt{suy10b}). At the same time, with so
many constraints placed on the model, Einstein ring lenses are
sensitive to perturbations from the smooth gravitational potential of
the lens galaxy \citep{koo05,veg09a}.  Satellite galaxies and dark
matter sub-haloes that orbit the main lensing galaxy can give rise to
these perturbations, and while these objects are often too faint to be
directly observed, gravitational lensing can be used to detect them
indirectly through their mass signatures.  In this way, careful
scrutiny of Einstein ring lenses (or those with extended gravitational
arcs) provides one of the best opportunities to reveal the presence
of, and constrain the properties of, extragalactic substructure
\citep{more09,vck10,veg10,veg12,suy10}.  This would serve as a direct
test of the $\Lambda$CDM numerical simulations that predict this
substructure \citep[e.g.,][]{die08,spr08}, and would thus advance our
understanding into the nature of galaxy formation and evolution.

For the reasons described above, we choose to study \lens\ using
high-resolution, NIR adaptive optics (AO) as part of the Strong
lensing at High Angular Resolution Program (SHARP). This new project
aims to obtain high resolution images of lens systems to study the
mass distributions of lensing galaxies between 0.3~$< z <$~1 to an
unprecedented detail with state-of-the-art imaging and lens modelling
techniques. In this first paper, we focus on a comparison between
imaging data taken with space- and ground-based observatories, in
particular, we determine the relative benefits for lens modelling. We
also present a new smooth mass model for our test system B1938+666
that is used to determine the properties of the lens galaxy and give a
robust estimate of the magnification of the NIR component of the
background galaxy.

This paper is organized as follows.  In Section \ref{sharp}, we
describe the wider goals of the SHARP survey since this is the first
paper in the series. In Section \ref{obs}, we briefly describe the new
and archival multi-wavelength imaging data for \lens, new IR
spectroscopy of the lensed source and the techniques used to reduce
them.  In Section \ref{results}, we present the results of lens
modelling.  We discuss our results in Section \ref{discussion}.
Finally, we summarize and conclude in Section \ref{conclusions}.
Measurements describing the relative fraction of substructure within
the \lens\ lensing galaxy are presented in a companion paper
\citep{veg12}.

Throughout this paper, we assume a cosmological model of $H_0 =
100\ h$ km\,s$^{-1}$ Mpc$^{-1}$, $\Omega_m = 0.3$, and
$\Omega_{\Lambda} = 0.7$.  All magnitudes presented in this work are
AB magnitudes.

\section{SHARP Rationale}
\label{sharp}

The image separations seen in galaxy-scale lenses are typically on the
order of an arcsecond.  Thus, observing these systems with high
angular resolution instruments can provide unambiguous estimates of
individual image positions and magnitudes, information that is
critical when measuring mass distributions with gravitational lensing.
Furthermore, any improvements in angular resolution compared to
traditional ground-based imaging will provide enhanced detectability
of faint objects with small angular extent, and improved sensitivity
to small-scale perturbations of the surface brightness of lensed
extended emission.  At optical and NIR wavelengths, there are
currently two techniques for obtaining the necessary data: space-based
imaging with \emph{HST} and ground-based AO imaging.  The SHARP survey
utilizes both approaches in order to build up a statistically
significant sample of lens systems for which deep high-resolution
imaging has been obtained.

\begin{figure*}
\begin{center}
\centerline{
\includegraphics[width=5.8cm]{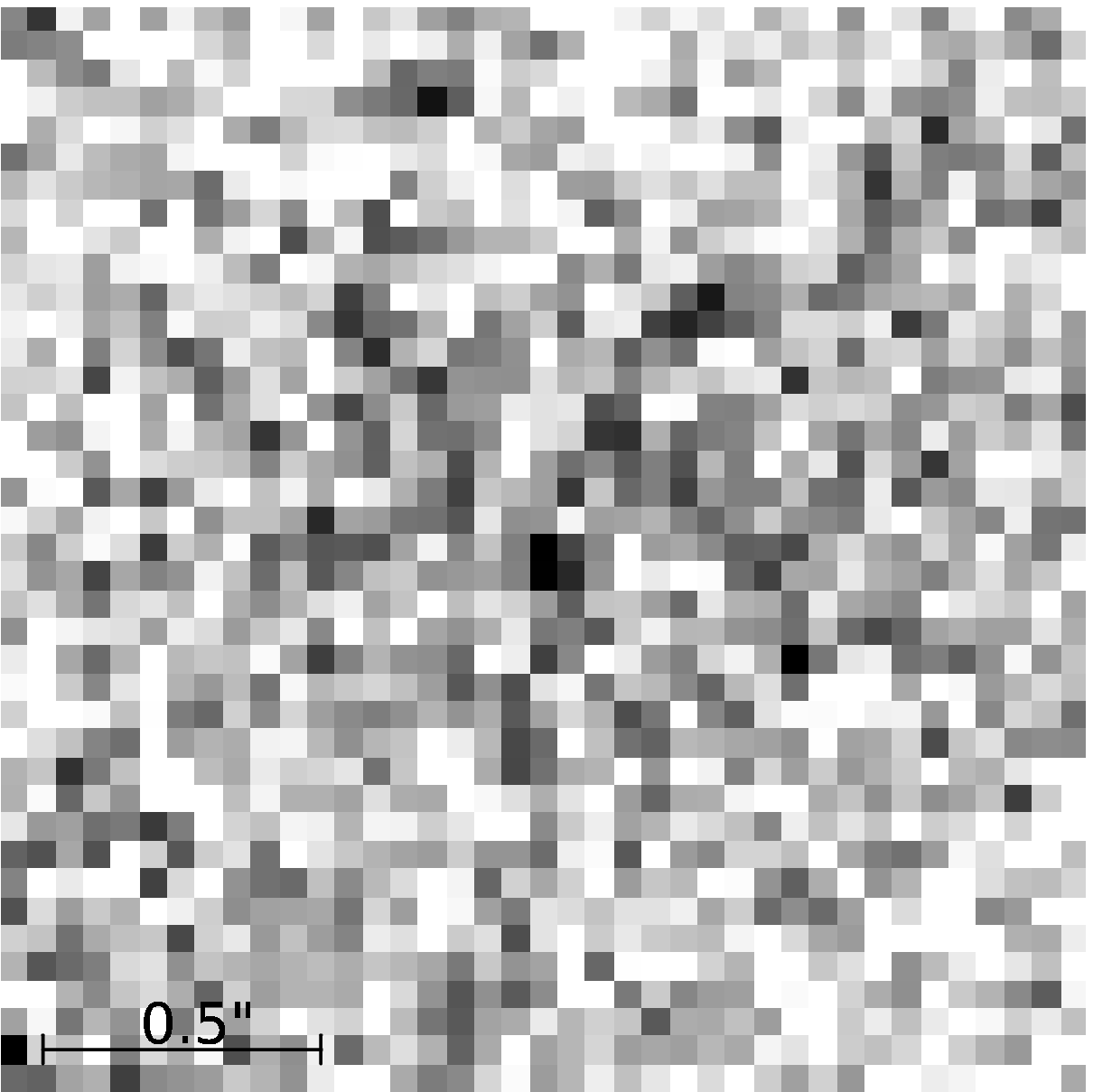}
\includegraphics[width=5.8cm]{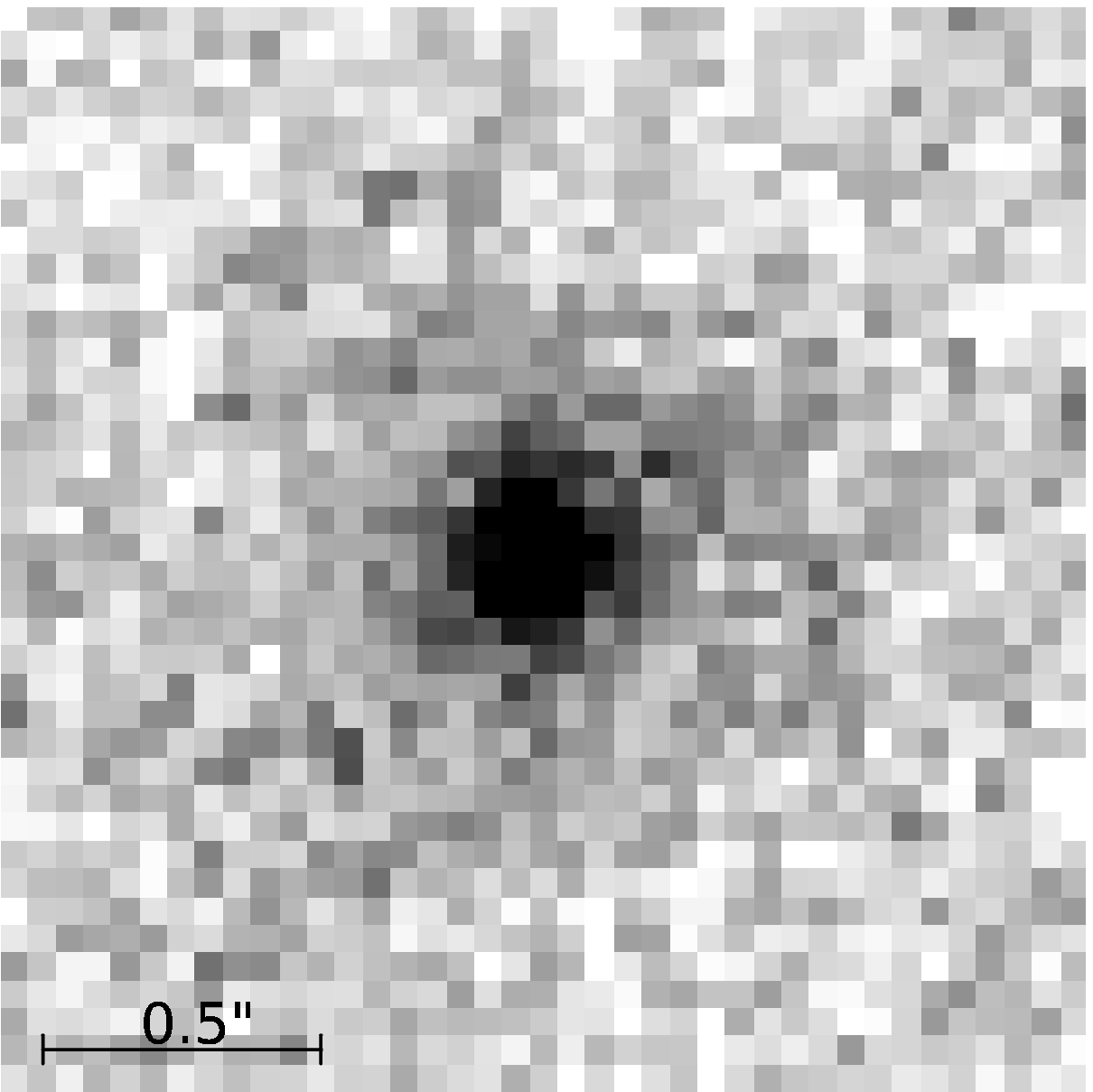}
\includegraphics[width=5.8cm]{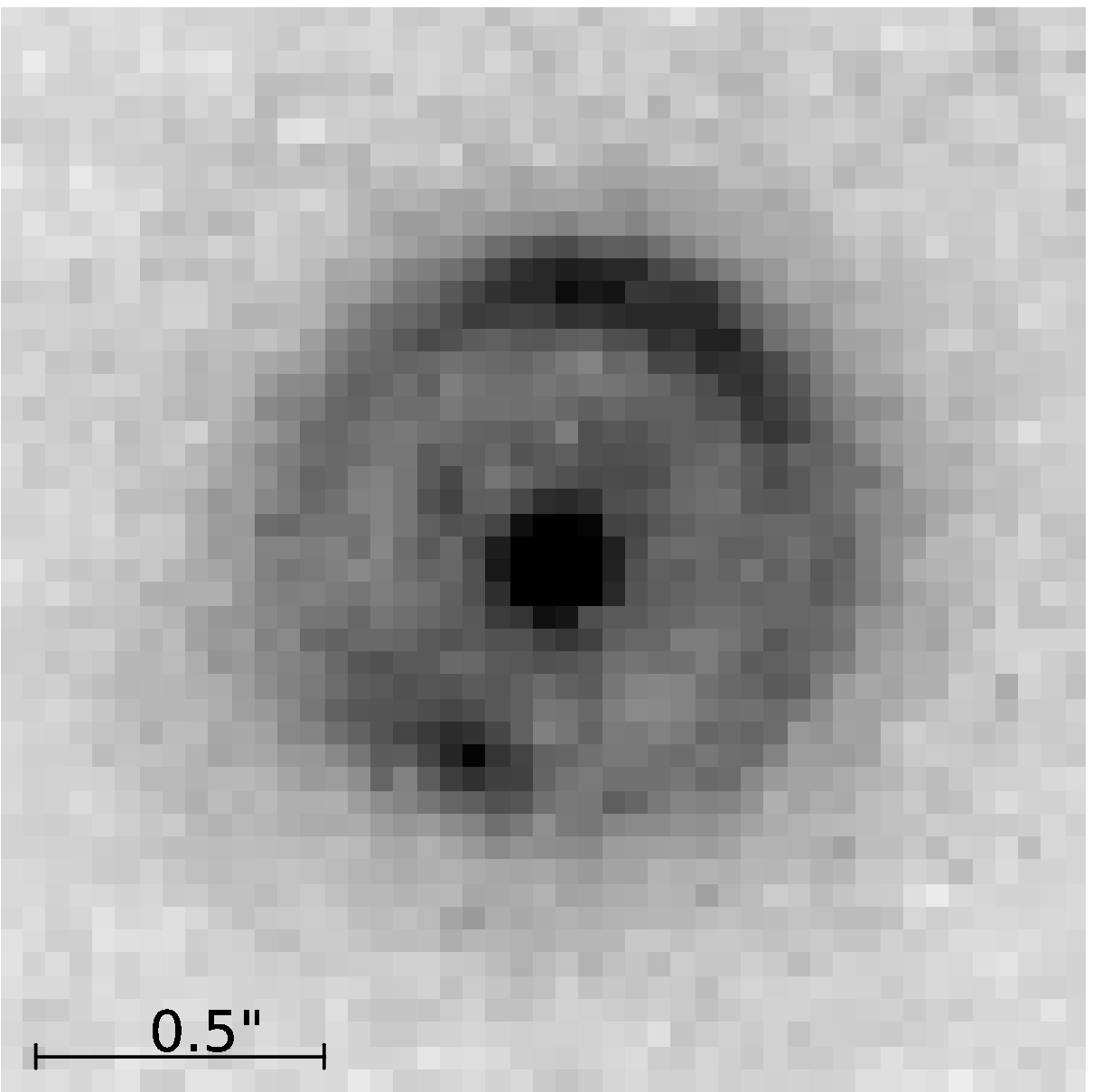}}
\centerline{
\includegraphics[width=5.8cm]{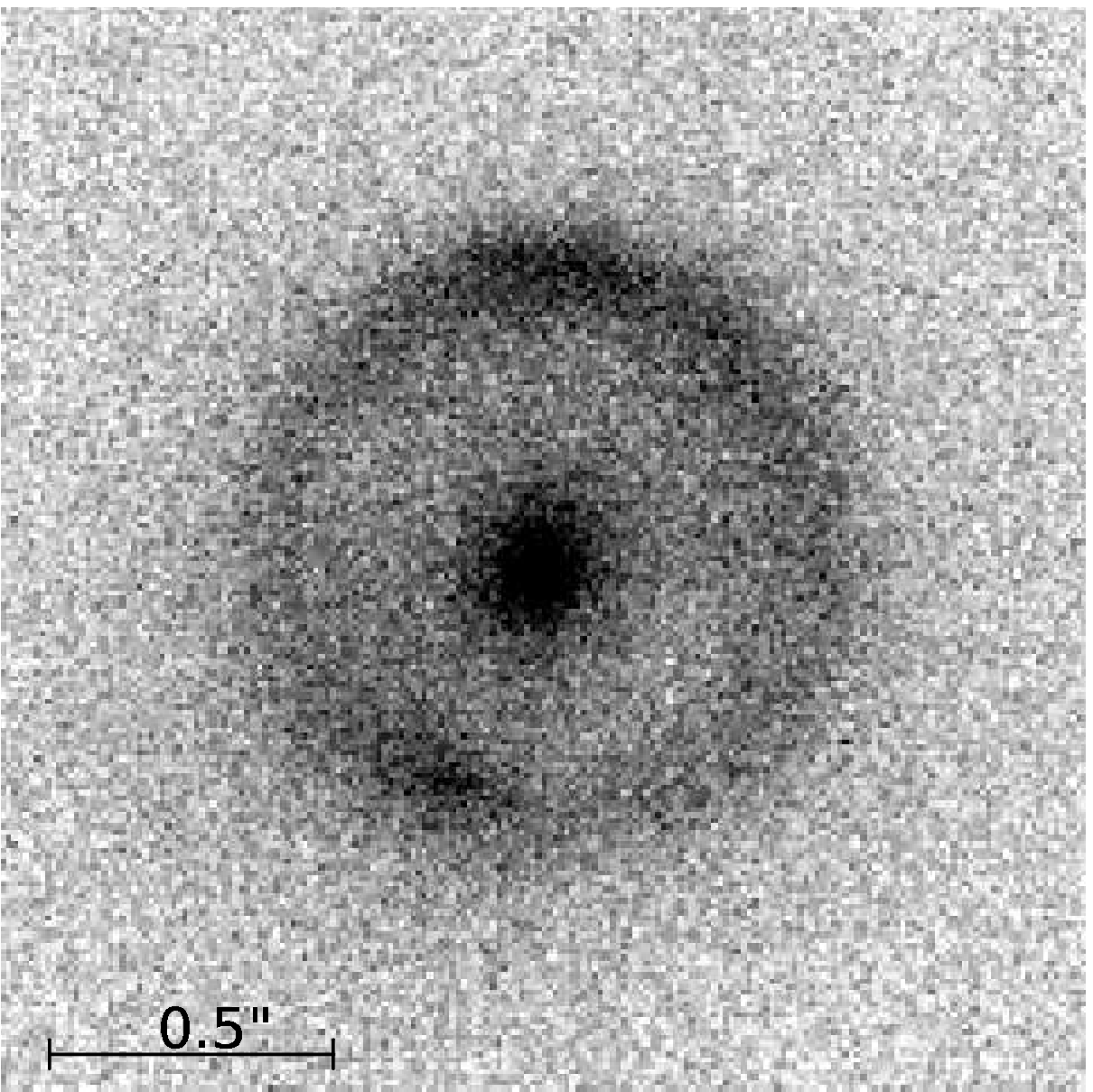}
\includegraphics[width=5.8cm]{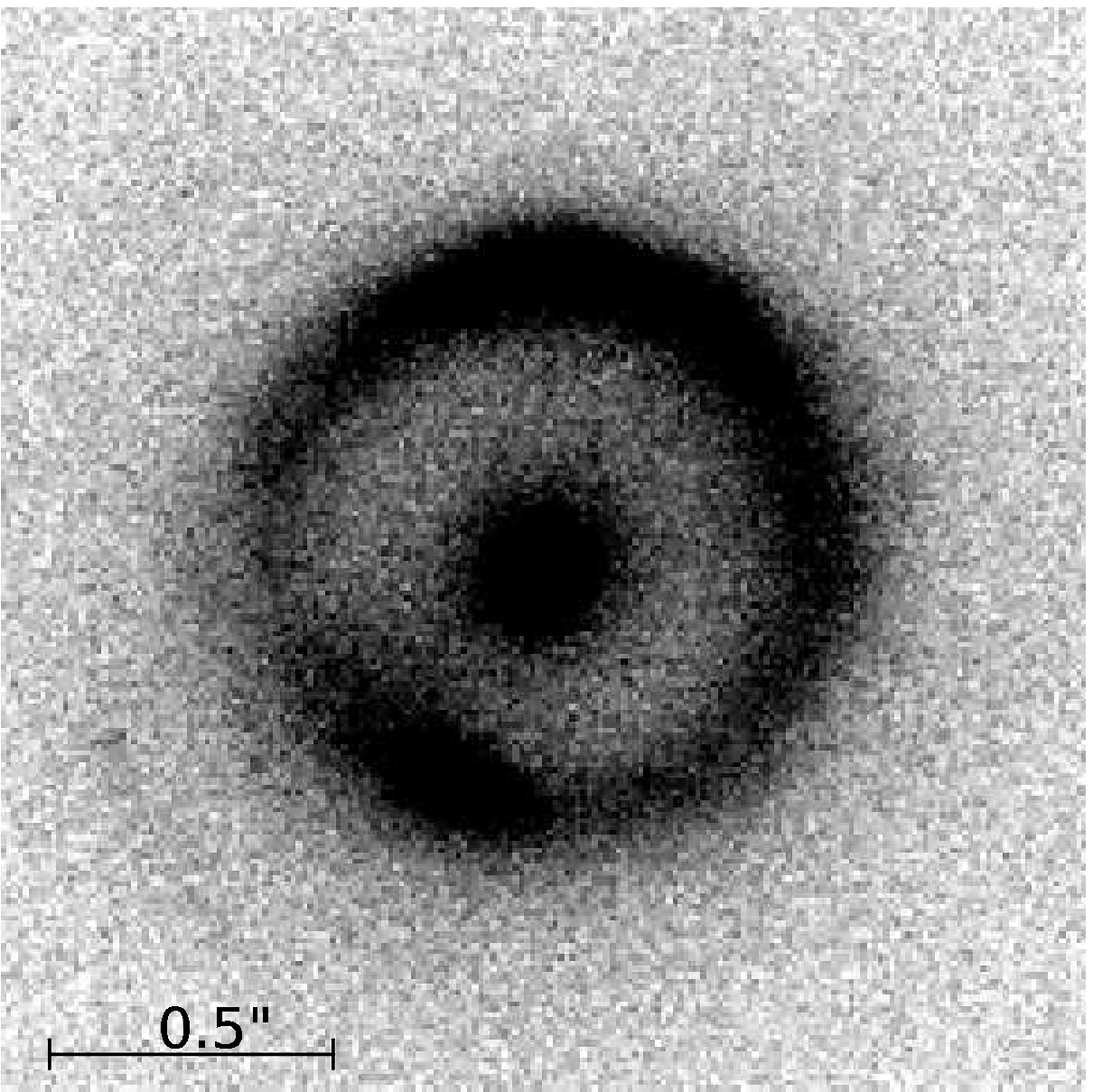}
\includegraphics[width=5.8cm]{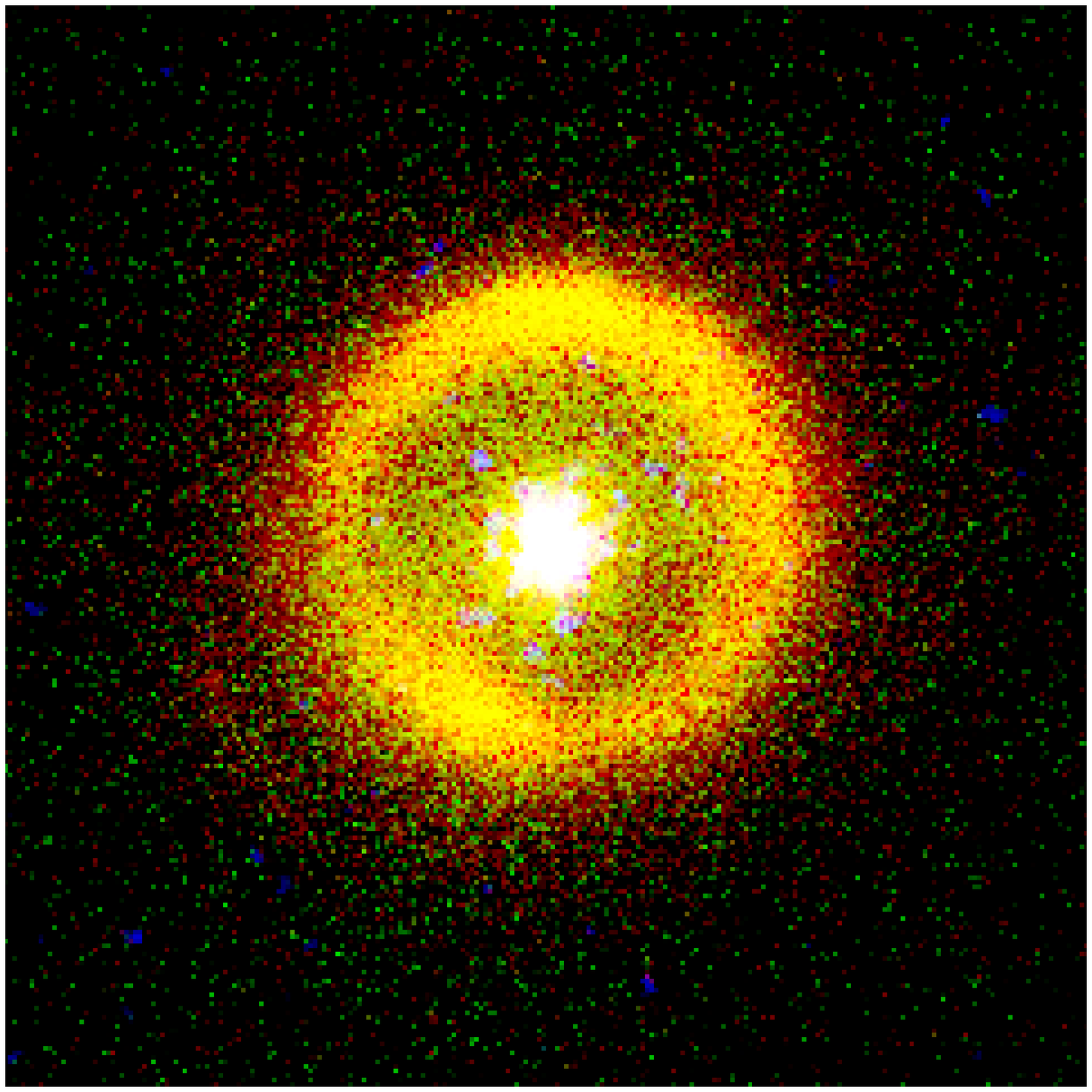}}
\end{center}
\caption{Multi-wavelength imaging of \lens.  Top row: \textit{HST}
  images taken using the F555W (V-Band), F814W (I-Band), and F160W
  (H-Band) filters, respectively from left to right. Bottom row:
  Keck-II Telescope LGS AO images showing the H-band (left panel) and
  K$^{\prime}$-band (middle panel) light. A colour composite image
  (right panel) was made by combining the \textit{HST} V- and I-band
  images with the LGS AO H- and K$^{\prime}$-band images. All
    images are oriented with North pointing up and East pointing to
    the left. The lensing galaxy and source galaxy are clearly seen
  in all of the near-infrared bands, whereas only the lensing galaxy
  can be seen in the optical bands.  Additionally, a second set of
  faint arcs can be seen on the eastern side of the K$^{\prime}$-band
  image (see Section \ref{jackpot} for further details).}
\label{fig:imgs}
\end{figure*}

A particular focus of the SHARP survey is to detect and measure the
mass of substructures associated with the lensing galaxies, without
regard to whether the substructures are luminous or dark.  Numerical
simulations of galaxy formation predict a large amount of substructure
for a galaxy-mass halo, with a mass fraction of $f_{\rm sub} =$
5--10~per cent of the total halo mass contained in substructures with
masses between 4~$\times $~10$^6 M_\odot$ and 4~$\times$~10$^9
M_\odot$ \citep{die08,spr08}.  Furthermore, the simulations converge
on a substructure mass function of $dN/dm \propto m^\alpha$, where
$\alpha = -$1.9~$\pm$~0.1 \citep[e.g.,][]{die07}. \citet{veg09b} have
shown that a Bayesian analysis of a sample of lenses that have been
surveyed for substructure down to some mass threshold can provide
meaningful constraints on $f_{\rm sub}$ and the slope of the mass
function, $\alpha$.  Given a sample of $\sim$30 lens systems, a mass
detection threshold of 3~$\times$~10$^8 M_\odot$ or better, and a
reasonable prior on $\alpha$, good constraints on $f_{\rm sub}$ are
obtained.  Note that even the non-detection of substructure down to
the mass limit provides useful information for constraining $f_{\rm
  sub}$ and $\alpha$, especially as the mass detection limit becomes
smaller.

We are therefore pursuing two complementary methods for detecting
substructure in lens systems, both of which utilize deep,
high-resolution imaging.  The first is to detect luminous satellites
directly, thus fixing the location of the substructure.  The
substructure can therefore be included in the mass model of the lens
and its mass can be determined \citep[e.g.,][]{more09}.  This method
is most effective for lens systems with either four lensed images or
extended lensed emission.  The other approach is to detect
substructures through their gravitational effects on extended lensed
emission \citep[e.g.,][]{veg10}.  Thus, the SHARP sample consists of
lens systems with quadruply imaged quasar lenses and lenses with
partial or complete Einstein rings.  Results on two four-image lenses
have been reported by \citet{mck07} and \citet{lag10b}.  Here we
report the results for B1938+666, which has an almost complete
Einstein ring.  A detailed description of the SHARP sample will be
presented by Fassnacht et al. (in preparation).

\section{Observations and Data Reduction}
\label{obs}

\subsection{Adaptive optics imaging}
\label{ao}
We observed the \lens\ system on UT 2010 June 29 and 30 with the NIRC2
camera on the Keck-II telescope, using the Laser Guide Star (LGS) AO
system.  We used the narrow camera, with a field-of-view of 10
arcsec~$\times$~10 arcsec, and a pixel scale of 0.01 arcsec.  The
tip-tilt correction was obtained through simultaneous observation of a
magnitude R = 15 star at a distance of 18 arcsec from the lens.  We
observed the system using both the H and K$^{\prime}$ filters.
Details of the observations are given in Table \ref{tbl:obs}.

\begin{table}
\caption{A summary of the B1938+666 optical and infrared imaging observations.}
\label{tbl:obs}
\begin{tabular}{ccccr}
\hline
Date        & Telescope  & Instrument  & Filter        & 
\multicolumn{1}{c}{$t_{\rm exp}$ (s)}\\
\hline
1997 Aug 13 & {\it HST}   & NICMOS/NIC1  & F160W       & 10800 \\
1999 Apr 24 & {\it HST}   & WFPC2        & F555W       &  2800 \\
1999 Apr 24 & {\it HST}   & WFPC2        & F814W       &  3000 \\
2010 Jun 29 & Keck II     & NIRC2 LGSAO  & K$^{\prime}$ & 14760 \\
2010 Jun 30 & Keck II     & NIRC2 LGSAO  & H           &  6600 \\
\hline
\end{tabular}
\end{table}

The data were reduced following the method presented in \citet{aug11}.
The final reduced images from the AO observations are shown in the
lower row of Fig.  \ref{fig:imgs}.  The images have three distinct
features: a bright, compact lensing galaxy in the centre of the image,
a high signal-to-noise ratio (SNR) Einstein ring, and a set of faint
arc structures outside of the ring on the east side.  Since we did not
observe any photometric standards during the data acquisition, we
calibrate our photometry by scaling the observed flux to match with
previously reported data.  The total (lens + ring) H-band flux is
matched to the NICMOS data presented by \citet{kin98} (F160W
$=$~19.4~$\pm$~0.3), while the K$^{\prime}$-band data are matched to
the NIRC value reported by \citet{rho96} (K$^{\prime}
=$~19.0~$\pm$~0.1).

\subsection{Archival {\it Hubble Space Telescope} imaging}
\label{hst}

The \lens\ system has previously been observed with the {\em HST}, at
NIR (GO 7255; PI Jackson) and optical (GO 7495; PI Falco) wavelengths.
The optical data were obtained as part of the CfA-Arizona Space
Telescope Lens Survey (CASTLES)
program\footnote{http://www.cfa.harvard.edu/castles/}.  These
observations are also summarized in Table~\ref{tbl:obs}.  The {\em
  HST} data were obtained from the archive and reduced using the
routines described in \citet{aug09}.  The final images are presented
in the upper row of Fig. \ref{fig:imgs} and, in the case of the NIR
imaging, show the same structure that was observed in the Keck AO
data.  In the F555W and F814W bands, the Einstein ring was not
detected.

\subsection{Near infrared spectroscopy}
\label{nirspec}

\begin{figure}
\begin{center}
\centerline{
\includegraphics[width=9.3cm]{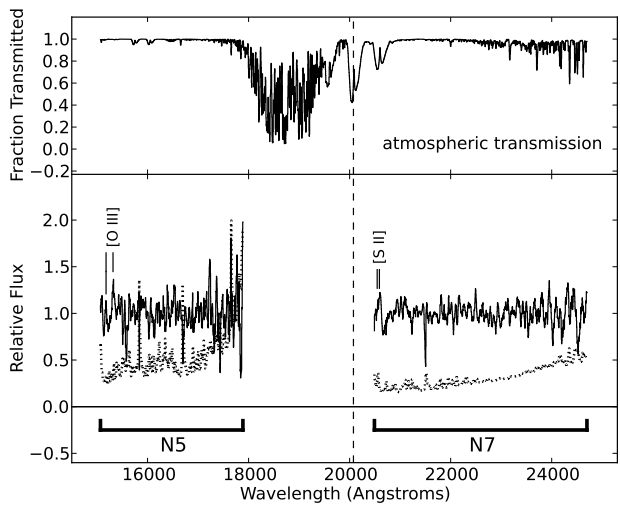}}
\end{center}
\caption{The NIRSPEC spectra of the combined emission from the lensing
  galaxy and background source (solid) and the 1$\sigma$ error
  spectrum (dotted). The spectra have been smoothed with a nine-pixel
  boxcar, with the points weighted by the inverse of their variances.
  The expected positions of the [O{\sc iii}] $\lambda\lambda$4960,
  5007 and [S\,{\sc ii}] $\lambda\lambda$6716, 6731 lines at the
  background source redshift of $z=$~2.059 \citep{rie11} are marked,
  while the vertical dashed line represents the expected position of
  H$\alpha$.  Additionally, the atmospheric transmission over the
  entire wavelength range of interest is shown in the upper panel.}
\label{fig:spec}
\end{figure}

The \lens\ system was observed with the near-infrared echelle
spectrograph (NIRSPEC; \citealt{mcl98}) on the Keck-II telescope on
the night of UT 2006 August 14.  The observing conditions were
marginal; scattered clouds throughout the night significantly altered
the sky transparency, which led to poor data quality.  The spectra
were obtained through the NIRSPEC-5 (roughly H-band) and NIRSPEC-7
(roughly K-band) filters.  Four exposures were taken in each band,
with each exposure consisting of three co-adds of 300~s.  Due to the
poor observing conditions, only two of the NIRSPEC-5 exposures were
usable.  All four NIRSPEC-7 exposures were used.  The spectra were
reduced with a Python-based pipeline, which consists of steps to
subtract the sky emission, reject cosmic rays, rectify the
two-dimensional spectra, wavelength calibrate, and extract the object
spectra.  The final spectra have pixel scales and spectral resolutions
of 3.56 \AA~pix$^{-1}$ and $\sim$1600 in NIRSPEC-5 and 4.08
\AA~pix$^{-1}$ and $\sim$1400 in NIRSPEC-7.

The source redshift was unknown at the time of the observations, and
both the H$\alpha$ and H$\beta$ lines fall outside the wavelength
range covered by the spectra.  However, given the source redshift of
$z_s =$~2.059 \citep{rie11}, the [O\,{\sc iii}] $\lambda\lambda$4960,
5007 emission features should fall in the NIRSPEC-5 spectrum and the
[S\,{\sc ii}] $\lambda\lambda$6716, 6731 features should fall in the
NIRSPEC-7 spectrum.  The spectra are shown in Fig.~\ref{fig:spec}.
They have been normalized and then smoothed by a 9-pixel moving
average, with each point being inverse-variance weighted.  Even with
the smoothing, no clear features are seen in the spectrum.  However,
the expected locations of the [O\,{\sc iii}] and [S\,{\sc ii}]
emission lines (marked in Fig.~\ref{fig:spec}) do correspond to weak
peaks in the spectrum.  Thus, although we would not have been able to
unambiguously measure the source redshift with the NIRSPEC spectra, we
can say that our data are consistent with the redshift measured by
\citet{rie11}.

\section{The lens model}
\label{results}

Strong gravitational lenses with extended-source structures are
frequently modelled by first determining and subtracting the surface
brightness distribution of the foreground galaxy
\citep[e.g.,][]{bol06}, although in cases where the background object
is very bright this can lead to an over-subtraction, where parts of
the strongly lensed features are fitted and removed along with the
foreground galaxy light \citep[e.g.][]{aug11}. At NIR wavelengths, the
(lensed) \lens\ source galaxy is roughly equal in brightness to the
foreground (lensing) galaxy, so the system presents a similar
modelling challenge. Therefore, we first fit a simply parametrized
lens model to the data in order to quantify and remove the foreground
galaxy light. We then fit a more detailed lens model to the residual
data, in order to precisely infer the properties of the mass
distribution.

\subsection{Surface brightness modelling}
\label{sb-model}

Modelling the surface brightness distribution for this system requires
that we also determine an approximate model for the lensing mass
distribution, in order to disentangle the foreground lens and
background source light.  To do this, we take an approach similar to
\citet{aug11} and employ an elliptical power law mass model
\citep[e.g.][]{bar98} to describe the mass distribution, while
including an external shear contribution.  The foreground and
background galaxy surface brightness distributions are modelled with
(possibly multiple) \citet{ser63} profiles. The models are fitted to
the data using an adaptive simulated annealing scheme, and each
potential model includes either a single surface brightness component
for both the lens and source galaxies, a single surface brightness
component for one galaxy and two components for the other, or two
surface brightness components for each galaxy.

The foreground galaxy is well-modelled with a single S{\' e}rsic
component, while the background source modelling strongly favours two
components. Although we find that there is significant covariance
between the structural properties (i.e., the S{\' e}rsic indices,
effective radii, and total magnitudes) of the foreground and
background components when fitting to the AO images, the {\it central}
surface brightness distributions are robustly segregated. For example,
the total magnitude of the foreground galaxy can change by 0.5
magnitudes if one or two components are used for the background
source, but the flux within the Einstein radius -- the radius of the
ring produced when a part of the lensed object sits directly behind
the lensing galaxy -- only varies by about 10~per cent ($\sim$0.1
magnitudes); the covariance between the total galaxy properties is
likely due to inadequate knowledge of the AO PSF. We therefore
restrict our discussion of the surface brightness properties to the
inferred luminosity within the Einstein radius and the position of the
source relative to the lens. The aperture magnitudes within the
Einstein radius are 21.0\,$\pm$\,0.15 and 20.6\,$\pm$\,0.10 in the H-
and K$^{\prime}$-bands, respectively.  Note that with lensing we
measure the total mass properties within the Einstein radius and we
are therefore able to draw robust conclusions about the relationship
between mass and light in the centre of the lensing galaxy, which we
discuss in Section \ref{lens}.

\subsection{Mass Modelling}
\label{model_A}

We model the Keck AO H- and K$^{\prime}$-band data sets along with the
{\it HST} NICMOS F160W-band data set using the adaptive and grid-based
Bayesian technique of \citet{veg09a}.  An extensive description of the
modelling procedure in the context of Bayesian evidence optimization
is provided there. In short, we model the system by assuming an
elliptical power-law mass distribution [$\rho(r)\propto r^{-\gamma}$]
for the lens galaxy, which we use to reconstruct the surface
brightness of the source galaxy (after first subtracting off the lens
galaxy light profile; see Section \ref{sb-model}) on a two-dimensional
grid.  This grid is adaptive in magnification and built by casting
pixels from the image plane back to the unlensed source plane.  Due to
the high resolution of the AO data, we only cast one point out of each
six-by-six pixel sub-grid back to the source plane, whereas for the
lower resolution \textit{HST} data, we use a smaller three-by-three
pixel sub-grid.  Note that the full pixel grid is used when the
reconstructed source surface brightness is cast forward to the image
plane and the residuals are calculated.

\begin{figure*}
\begin{center}
\subfigure[]{\centering\includegraphics[width=0.48\hsize]{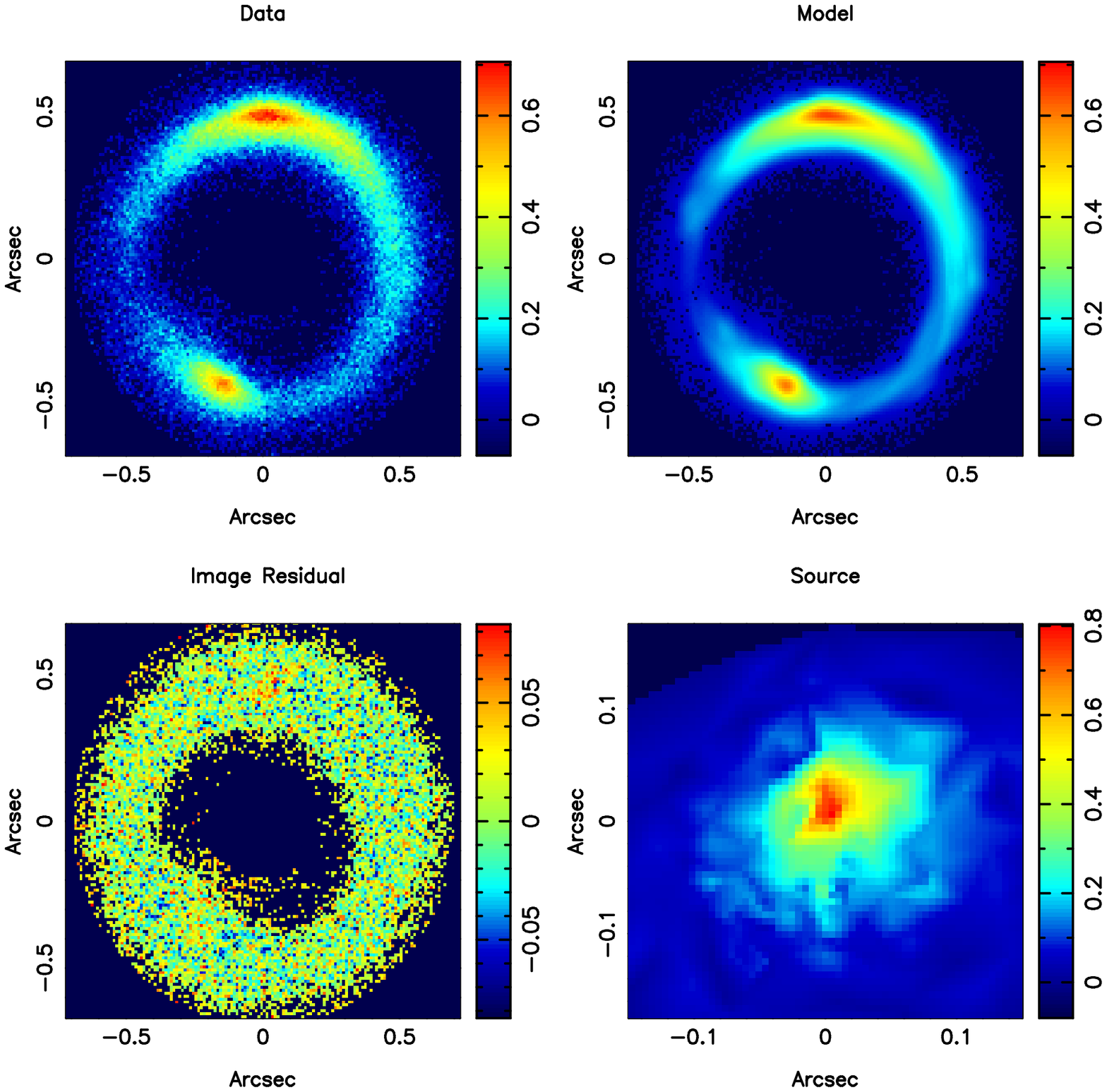}
\label{fig:model-ao}
}
\hfill
\subfigure[]{\centering\includegraphics[width=0.48\hsize]{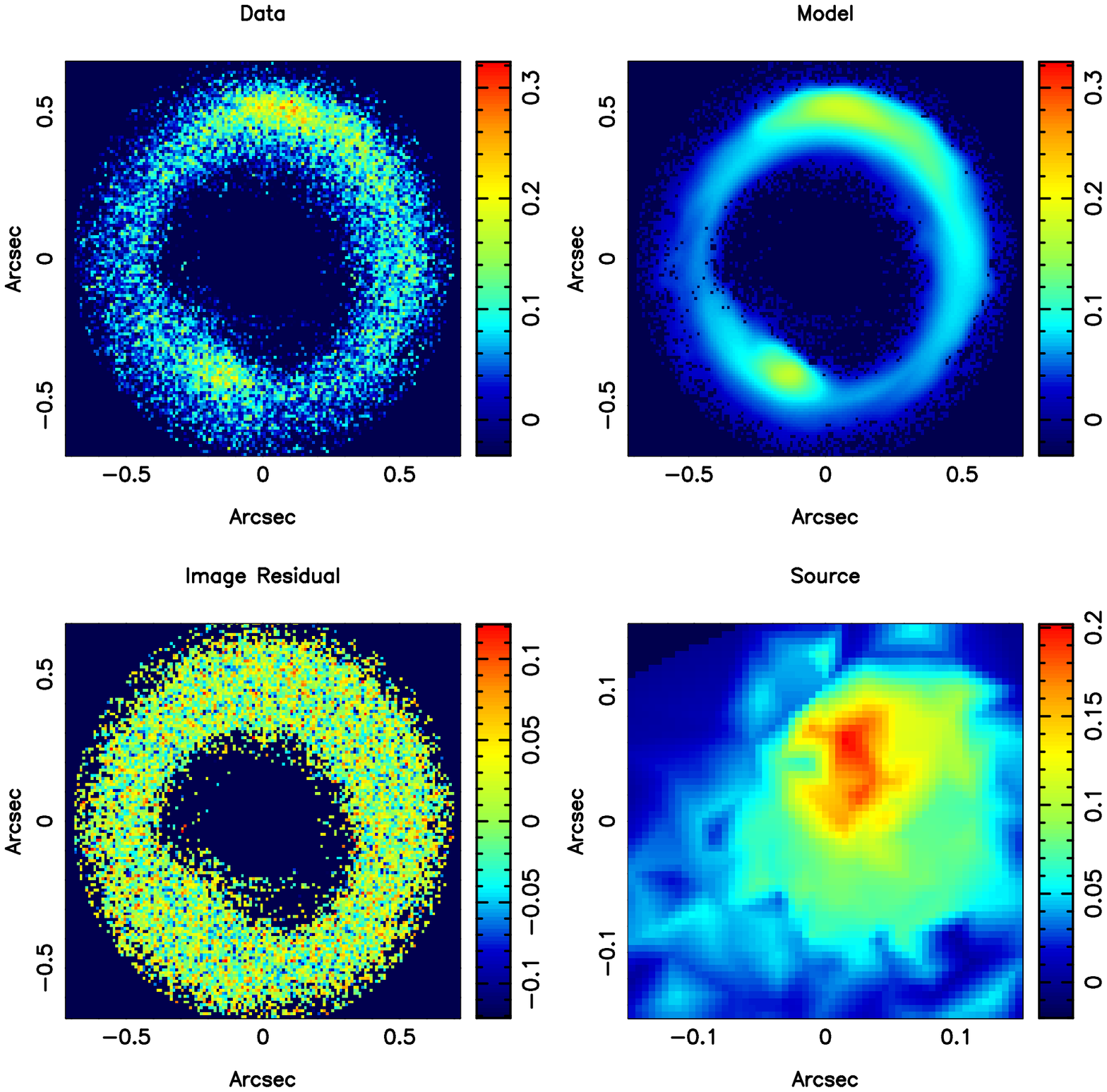}
\label{fig:model-ao2}
}
\hfill
\subfigure[]{\centering\includegraphics[width=0.48\hsize]{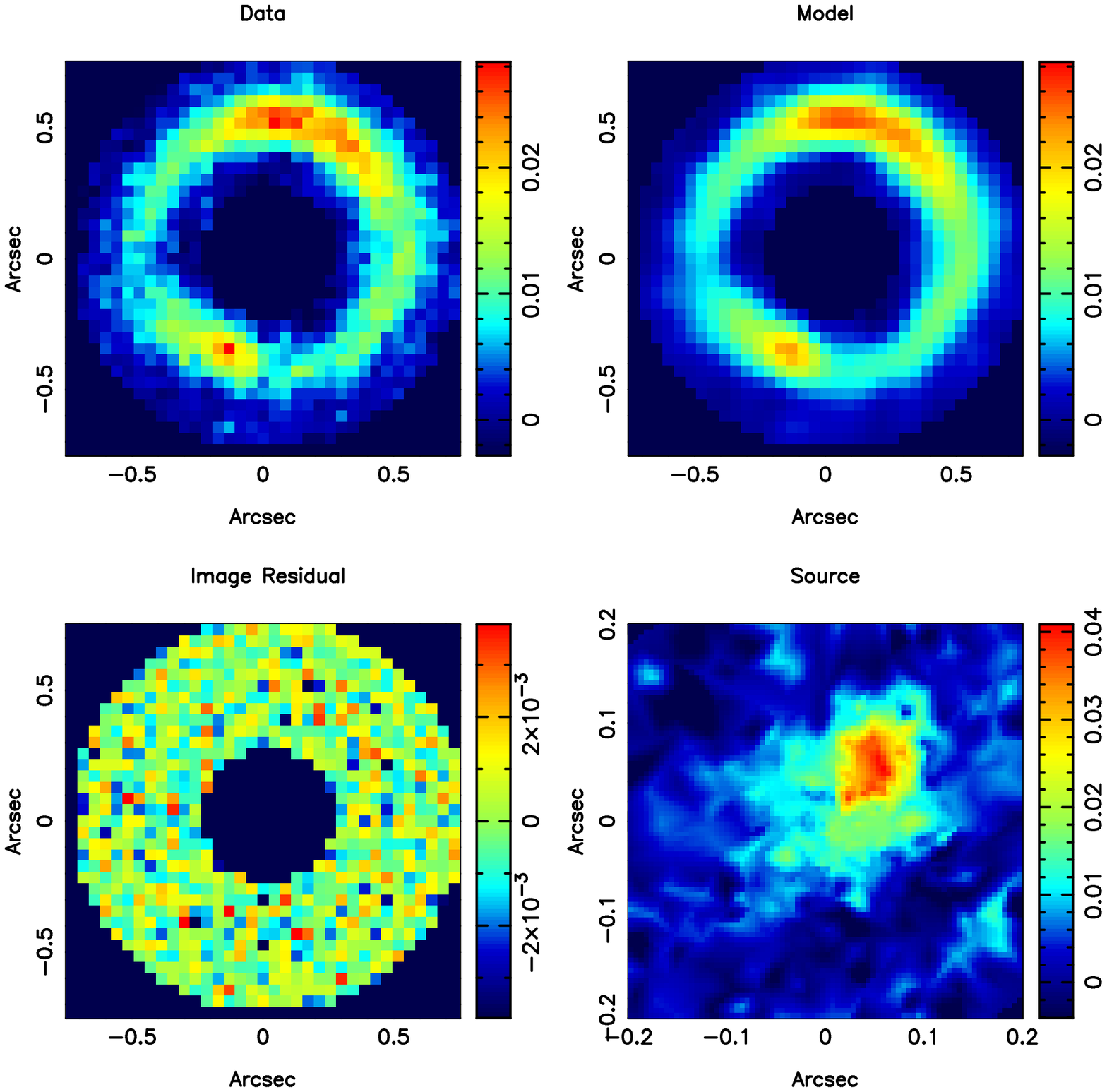}
\label{fig:model-nic}
}
\caption{The gravitational lens mass model of the \lens\ Einstein
  ring, using the adaptive grid-based method of \citet{veg09a}. The
  results are for the three independent data sets: (a) the AO
  K$^\prime$-band, (b) the AO H-band and (c) the {\it HST} NICMOS
  H-band. For each plot, we show the measured surface brightness
  distribution of the Einstein ring (top left panel), the best-fit
  smooth lens model reconstruction (top right panel), the residual
  image (bottom left panel) and the reconstructed unlensed image of
  the background source galaxy (bottom right panel).  We note that the
  source reconstruction grids are not registered.  Therefore, taking
  into account registration offsets, as well as small offsets due to
  PSF convolution, we find that the source positions are consistent
  with being coincident in all data sets.}
\end{center}
\end{figure*}

\begin{figure*}
   \begin{center} 
      \includegraphics[width=18cm]{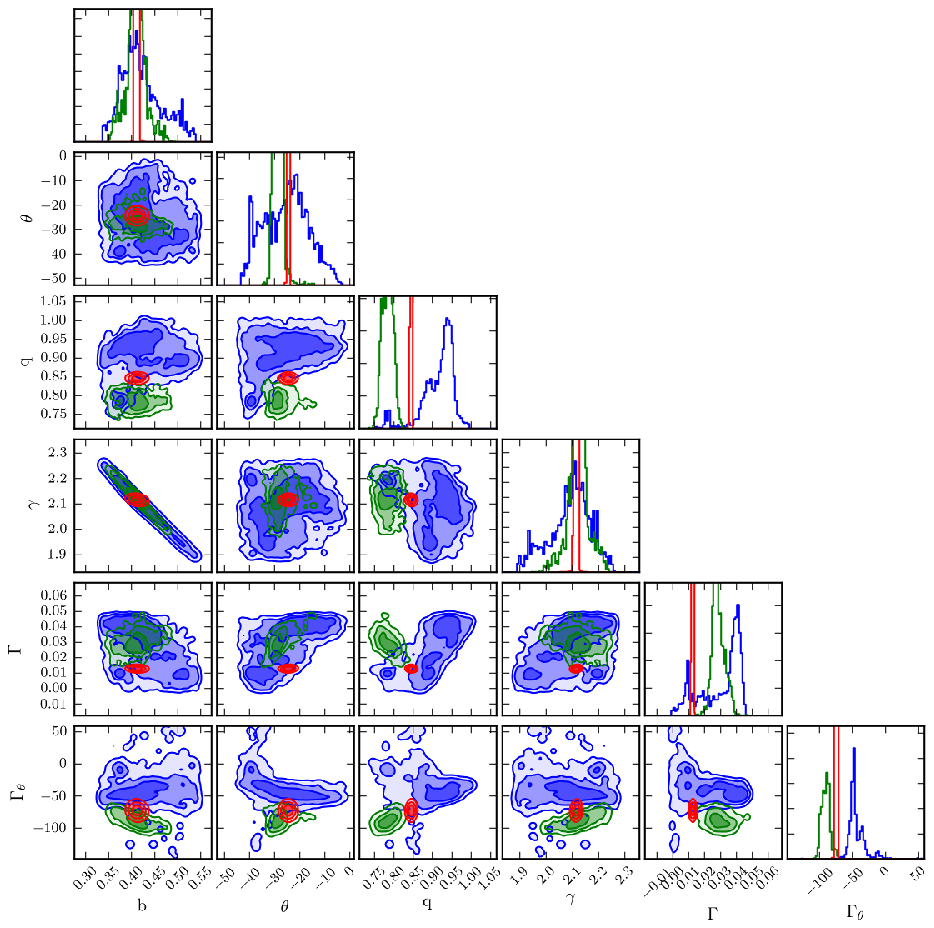}
            \caption{The marginalized posterior
              probability distributions of the lens model parameters,
              as measured by a nested sampling analysis.  Each grid
              shows a different marginalization: the 2-dimensional
              contours represent the distribution between two model
              parameters (specified by the row and column), while the
              1-dimensional histograms at the top of each column
              represent the distribution of a single parameter.  The
              AO K$^{\prime}$-band, AO H-band, and NICMOS data are
              represented by red, green, and blue contours,
              respectively. The maximum-likelihood model values for
              each parameter, and their uncertainties, are presented
              in Table \ref{tbl:modeling_results}.}
      \label{fig:posterior} 
    \end{center}     
 \end{figure*}

The best lens models (as determined from maximum-likelihood analysis)
for the K$^{\prime}$-band, H-band, and NICMOS data sets are presented
in Figs. \ref{fig:model-ao}, \ref{fig:model-ao2} and
\ref{fig:model-nic}, respectively, while the individual parameters of
these models are listed in Table \ref{tbl:modeling_results}.  In
addition to the maximum-likelihood values, we also derive mean values
and confidence intervals for each parameter's marginalized posterior
probability distribution function by exploring the Bayesian evidence
in the full multidimensional parameter space (see Table
\ref{tbl:modeling_results} for the results).  One- and two-dimensional
slices of these marginalized posterior probability distributions are
shown in Fig.~\ref{fig:posterior}.  Using Multinest v2.7
\citep{fer08}, we are able to integrate over this posterior
probability and obtain the marginalized Bayesian evidence, which is
the probability of the data given the model family.

Incorporating a robust PSF model can be challenging for AO data, since
the PSF varies rapidly over time. This is a particular concern for the
\lens\ AO data sets as we are unable to observe the lens system and a
PSF star simultaneously.  Therefore, we carry out the modelling using
different PSF stars taken at different times, leading to a different
``best'' model for each PSF.  As the arc is sufficiently extended and
the dynamic range is relatively low, we find that the lens modelling
of \lens\ is not significantly affected by the choice of PSF (see
\citealt{veg12} for details).  However, the Bayesian evidence allows
one to choose the best PSF model objectively. For example, the PSF
used for the model labelled $\rm M_K$ in
Table~\ref{tbl:modeling_results} yields the largest evidence value for
the K$^\prime$-band data.

The best models for each data set are consistent, as can be seen by
comparing the maximum-likelihood parameters given in Table
\ref{tbl:modeling_results}.  Many of the marginalized posterior
probability distributions are also in agreement
(Fig.~\ref{fig:posterior}), though we do note that there are
discrepancies in the axis ratios and external shear parameters,
especially between the NICMOS and H-band models (see Section
\ref{comparison}).  However, the precision with which the lens
parameters can be recovered is significantly higher for the Keck AO
data, with the highest precision obtained for the
K$^{\prime}$-Band. This clearly shows that the higher resolution
provided by the AO imaging allows for a more precise lens modelling,
when compared to the {\it HST} imaging, despite the lower
signal-to-noise ratio (SNR).

Finally, it is interesting to note that all of the models using the
adaptive and grid-based Bayesian technique lead to a total mass
centroid position and flattening that are consistent with those of the
parametric mass modelling presented in Section~\ref{sb-model}. Also,
all of the models have a total density profile that is very nearly
isothermal (i.e., $\gamma =$~2) out to the Einstein radius of the
system.

\begin{table*}
\caption{Individual parameters of the grid-based reconstruction lens
  models.  The first row of a given model (specified by the first
  column) represents the maximum-likelihood model solution, while the
  second row gives the mean values for each parameter's posterior
  probability distribution.  The third and fourth rows show,
  respectively, the 68 and 95 per cent confidence intervals of the
  means.  The model parameters themselves are described as follows.
  $b$ is the model lens strength (not the Einstein radius as defined
  for a SIS mass model) in arcseconds.  $\theta$ is the position angle
  of the mass distribution of the lensing galaxy (in degrees east of
  north) while $q$ is the axis ratio.  $\gamma$ is the power-law slope
  of the mass profile.  $\Gamma$ and $\Gamma_{\theta}$ are,
  respectively, the magnitude and position angle (in degrees east of
  north) of an external shear source.  Additionally, the global
  Bayesian evidence of the model is presented in the final column.}

         \begin{tabular}{lccccccc}
       \hline

 Model  &$b$  &$\theta$  &$q$  &$\gamma$  &$\Gamma$  &$\Gamma_{\theta}$  &Evidence \\ 
 \hline			
$\rm{M_K}$  &0.452   &$-$22.4   &0.853   &2.05   &0.014   &$-$77.1 &        \\
            &0.413   &$-$24.3   &0.846   &2.12   &0.013   &$-$72.7  &48806.4 \\
& $[0.410, 0.415]$ & $[-24.5, -24.1]$ & $[0.845, 0.848]$ & $[2.11, 2.12]$ & $[0.012, 0.013]$ & $[-73.2, -72.2]$\\
& $[0.408, 0.418]$ & $[-24.6, -23.9]$ & $[0.843, 0.849]$ & $[2.10, 2.13]$ & $[0.011, 0.014]$ & $[-74.8, -71.6]$\\
$\rm{M_H}$  &0.447   &$-$22.3  &0.853   &2.05   &0.019   &$-$78.5 &        \\     
            &0.410   &$-$28.5  &0.786   &2.12   &0.029   &$-$89.1  & 46280.0\\
& $[0.397, 0.426]$ & $[-29.8, -27.3]$ & $[0.770, 0.801]$ & $[2.09, 2.15]$ & $[0.025, 0.032]$ & $[-97.9, -84.1]$\\
& $[0.360, 0.461]$ & $[-31.7, -25.9]$ & $[0.761, 0.811]$ & $[2.03, 2.22]$ & $[0.022, 0.038]$ & $[-101.0, -73.4]$\\
$\rm{M_{HST}}$   &0.439   &$-$23.0   &0.856   &2.07   &0.016   &$-$74.1  &        \\
                &0.424   &$-$25.0   &0.918   &2.09   &0.031   &$-$42.4  & 7027.16 \\
& $[0.372, 0.454]$ & $[-38.7, -16.8]$ & $[0.890, 0.957]$ & $[2.04, 2.20]$ & $[0.009, 0.044]$ & $[-53.8, -32.8]$\\
& $[0.355, 0.520]$ & $[-39.8, -8.0]$  & $[0.777, 0.988]$ & $[1.93, 2.23]$ & $[0.006, 0.046]$ & $[-59.0, -6.6]$\\
 \hline
\end{tabular}
\label{tbl:modeling_results}
\\
\end{table*}

\section{Discussion}
\label{discussion}

We now discuss our results. First, we give a comparison of modelling
the gravitational lens system using high angular resolution data from
space- and ground-based telescopes. We then discuss the properties of
the lensing galaxy and the magnification of the background source that
we infer from our model. Finally, we investigate the possibility of
there being a second gravitationally lensed source for this system.

\subsection{Comparing \emph{HST} and AO data sets for gravitational lens modelling}
\label{comparison}

The precision and accuracy of mass model constraints from extended
sources on strong gravitational lens galaxies is set by three
conditions: (i) the number of independent resolution elements across
the lensed images (and their multiplicity), (ii) the average SNR of
the lensed-image surface brightness distribution and (iii) the level
of surface brightness structure in the lensed source.  While the first
two conditions set the level of information contained in the lensed
images and how well, in principle, the mass model of the lens galaxy
can be constrained, the third condition determines the level of
covariance (i.e., degeneracy) in the mass model, where more structured
sources in general lead to a lower level of covariance between model
parameters.  One of the goals of this paper has been to illustrate
points (i) and (ii) by comparing the results of high SNR {\it HST}
F160W-band data with both lower resolution and pixel-sampling, and
lower SNR Keck AO imaging data with higher resolution and sampling. We
implicitly assume that the intrinsic source structure in the F160W-,
H-, and K$^{\prime}$-bands is very much correlated.

Despite the fact that the results of the modelling of the three very
different data sets are promisingly similar and the errors are small,
we do note that there are differences. In particular, there are
discrepancies between the axis ratio and position angle (PA) of the
lens mass distribution, and the external shear strength and PA (see
Fig. \ref{fig:posterior}).  Some tension between these quantities can
be expected, as they are strongly covariant (i.e., the external shear
mimics the flattening of the lens potential/mass-distribution) and
lack of information combined with errors on the data can lead to
biases in the maximum-likelihood solutions and the posterior
probability distributions of individual parameters.  This is
especially true for \lens, where a nearly circular mass distribution
and largely featureless Einstein ring conspire to add uncertainty to
the lens model.  Specifically, the circularly symmetric mass
distribution (coupled with nearly coaxial foreground and background
galaxies) makes it difficult to constrain the lensing galaxy's mass
slope -- increasing the covariances between other parameters -- while
the smooth light distribution lacks the contrast needed to
differentiate between small-scale variations on the model.  Thus, even
moderate variations in parameter space can leave the lensed light
distribution relatively unchanged, allowing significantly different
models to fit the data equally well.

Looking at the marginalized probability distributions in
Fig. \ref{fig:posterior} it is obvious that there is a strong
covariance between the lens strength ($b$) and mass slope ($\gamma$),
especially in the H-band and NICMOS models.  This is why model
discrepancies are the strongest between these data sets.  However, we
note that these degeneracies, and the parameter uncertainties in
general, are much smaller from the higher-resolution (but lower SNR)
Keck AO data. This implies that at least for some systems,
ground-based AO data can ``out-perform'' space-based data in precision
-- and most likely also in accuracy -- due to the better sampled lens
image structure.  Of course, this does not suggest that higher SNR
data is unnecessary: of the two AO-based models, the K$^{\prime}$-band
data (with its relatively higher SNR) provides much tighter
constraints on individual parameters than the H-band model, and
indeed, even breaks the mass-slope degeneracy found in the H-band and
NICMOS models.  Rather, the comparison simply shows that, between SNR
and image resolution, resolution seems to dominate any limitations on
model precision.

One has to be careful, though, when attempting to generalize this
conclusion to other lens systems, because the result depends
critically on the complex interplay between SNR, spatial resolution,
and source structure. It could very well be that lens systems with
highly-structured sources (e.g., space-based B- or U-band data, where
star formation could be strong and the sources could be more
structured) could out-perform the higher resolution K$^{\prime}$-band
AO data. This interplay is currently under study.  In cases where
there is similar data, and a bright star is available, however, our
results show that ground-based AO data can perform significantly
better than their space-based {\it HST} data sets in constraining lens
models. As we show in the companion paper, the higher resolution
provided by AO imaging was also very powerful in the discovery of a
low-mass substructure in this lens system \citep{veg12}.

\subsection{The foreground lensing galaxy}
\label{lens}

The \lens\ lensing galaxy is clearly seen in the optical and NIR data
sets (Fig. \ref{fig:imgs}), and its surface brightness distribution is
well-modelled by a single elliptical S\'{e}rsic profile (see Section
\ref{sb-model}). Excluding the HST V-band photometry, we find the
galaxy to have I$-$H and H$-$K$^{\prime}$ colours of 1.9 and 0.4
magnitudes, respectively, after Galactic reddening corrections. We
model these colours with the \citet{bc03} stellar population code and,
assuming a Solar metallicity and no dust, find consistency with an old
($\ga$~4~Gyr) stellar population. The foreground galaxy's isophotal
regularity, colours, and absorption line spectrum measured by
\citet{ton00} suggest that the lens has an early-type morphology; this
is consistent with the previous analysis of \citet{kin98} based only
on the NICMOS F160W-band imaging.

We use the \citet{bc03} 4~Gyr stellar population model to compute the
V-band absolute magnitude and to estimate the stellar mass within the
Einstein radius. The luminosity is found to be $L_{\rm V}
=$~3~$\times$~10$^{10}~L_\odot$ and the stellar template has a V-band
stellar mass-to-light ratio of 1.54, implying a stellar mass within
the Einstein radius of $M_* =$~4.7~$\times$~10$^{10}~M_\odot$ if a
Chabrier initial mass function (IMF) is assumed. We can compare this
to the \emph{total} mass within the Einstein radius from lensing, and
we find a dark-matter mass fraction of $M_{\rm dark}/M_{\rm lens}
=$~0.55 within the Einstein radius.  If we instead use a Salpeter IMF
to describe the stellar population, the stellar mass within the
Einstein radius becomes $M_* =$~ 8.2~$\times$~10$^{10}~M_\odot$,
leading to a dark-matter mass fraction of $M_{\rm dark}/M_{\rm lens}
=$~0.2.

Ideally, we would like to compute the stellar-to-total mass fraction
within an aperture physically associated with the lensing galaxy.
Therefore, we use two methods to estimate the effective radius from
the more robustly measured lensing data. First, we employ the
relationship between the power-law density slope and effective radius
found by \citet{aug10} and correct for early-type galaxy growth rates
\citep[e.g.,][]{new10}. There are several significant caveats,
including evolution of the power-law slope
\citep[e.g.,][]{ruff11,bol12} and large uncertainties on these
relationships; nevertheless, we find that the effective radius is
consistent with being the same size as the Einstein radius. Motivated
by this, we check if the assumption that the effective radius of
0.45~arcsec (i.e., equivalent to the Einstein radius) is consistent
with measurements of the fundamental plane.  We use the velocity
dispersion inferred from the lensing model as a proxy for the stellar
velocity dispersion and apply a passive evolution correction to the
luminosity as determined by the \citet{bc03} 4~Gyr template. We find
that, again correcting for size evolution, our assumption that the
effective radius and Einstein radius are co-incident yields
consistency with the fundamental plane presented by \citet{aug10}.

\subsection{The magnification of the NIR light from the background galaxy}
\label{source}

The gravitational lens models also provide an estimate of the source
surface brightness distribution, after correcting for the lensing
effect. From the pixelated reconstruction, we find that the source is
composed of two components at NIR wavelengths (in agreement with
the parametrized source distribution discussed in Section
\ref{sb-model}). One component is a high surface brightness region
with a projected size of $\sim$~0.8 kpc, the second component is more
extended, up to $\sim$1.6~kpc in projected size, and has a lower
surface brightness. We defer any detailed discussion about the nature
of the NIR source reconstruction -- especially with respect to the AGN
and molecular gas components traced by the emission at radio
wavelengths -- to a follow-up paper.

To estimate the total magnification of the NIR source from the
gravitational lens models, we compare the model emission found in the
lensing plane to that found in the source plane. We use the {\it HST}
NICMOS data set for this calculation because these data have the most
robust measurement of the extended light distribution of the source
galaxy; the AO data sets will have some light artificially pushed into
an extended envelope (thus leading to a greater chance of confusion
between source and lens galaxy light) due to the fact that AO Strehl
ratios are less than unity.  Additionally, the brightness of the sky
background at NIR wavelengths can hide faint, extended emission at the
outer edge of the source galaxy, leading to a biased magnification
estimate.

Overall, we find that the total magnification factor of the NIR
emission is $\sim$~13.  We note that a magnification of 176 has been
previously reported for B1938+666 from a simple point-source model
(see \citealt{bar02} for some details). This larger magnification was
used by \citet{rie11} for their analysis of the molecular gas
properties of B1938+666, and they found that the CO (3--2) line
intensity for B1938+666, and hence the molecular gas mass, was about
an order of magnitude lower than for other quasars at a similar
epoch. This large discrepancy is almost certainly due to the
magnification of the gas emission being over-estimated. However,
although we find a smaller magnification for the NIR emission region
for this source, some caution should be taken when applying this value
to other wavebands, particularly if the emission has a different size,
or position with respect to the lensing caustics produced by the lens;
only spatially resolved imaging and a good lens model can give a
robust estimate of the magnification for any particular emission
region of a gravitationally lensed source.

\subsection{Luminous Substructure}

\begin{figure}
\begin{center}
\includegraphics[width=7cm]{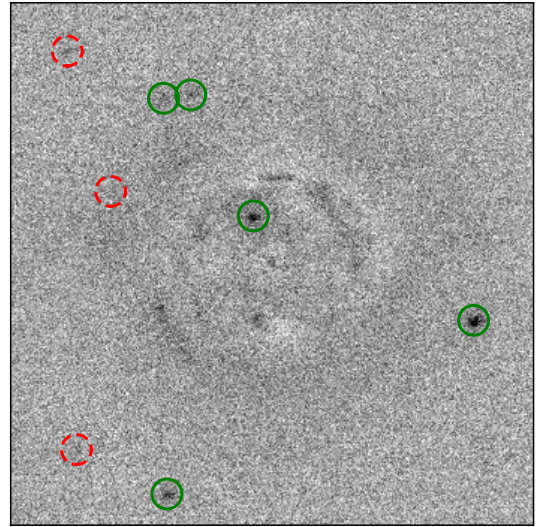}
\caption{Residual image of the \lens\ system with simulated point
  source objects added, which we use to estimate a detection limit for
  luminous substructure.  The point sources detected by SExtractor are
  represented by green, solid-line circles, while those that are
  missed are represented by red, dashed-line circles.  There is a
  distinct difference in noise level between the regions inside and
  outside of the Einstein ring, suggesting that the substructure
  limiting magnitude should be brighter closer to the lens.}
\label{fig:psim}
\end{center}
\end{figure}

In \citet{veg12} we presented the detection of a low-mass substructure
in the B1938+666 system from the distorting effect the substructure
has on the lensed arc. We do not find any evidence of luminous, and
probably more massive, substructure in the residual images for the
system, which is consistent with the \citet{veg12} analysis. To
determine an upper limit on the brightness of any luminous
substructure, we add a series of simulated point sources between
24.5~$\le m_{K^{\prime}} \le$~28 to an image of the lens system from
which both the lens galaxy and the Einstein ring emission have been
subtracted.  For our point source model, we choose the empirical PSF
star used in the lens modelling, as this would accurately represent
how a point source object would appear in the image.  One example
simulation is presented in Fig.  \ref{fig:psim}.  We choose a point
source model over other models -- such as Gaussian or S\'{e}rsic
profiles -- because we have no inherent knowledge about the
morphologies of these objects, and because the typical small size of
satellites relative to a parent galaxy, coupled with the high redshift
of the \lens\ lensing galaxy, suggests that these objects would be
likely unresolved in our data.  Additionally, since a point source
object has a higher surface brightness than an extended object of the
same magnitude, a point-source limiting magnitude represents a robust
limit to the lowest luminosity that can possibly be observed, placing
a hard limit on the luminous properties of any substructure that is
independent of morphology.

\begin{figure}
\begin{center}
\includegraphics[width=8cm]{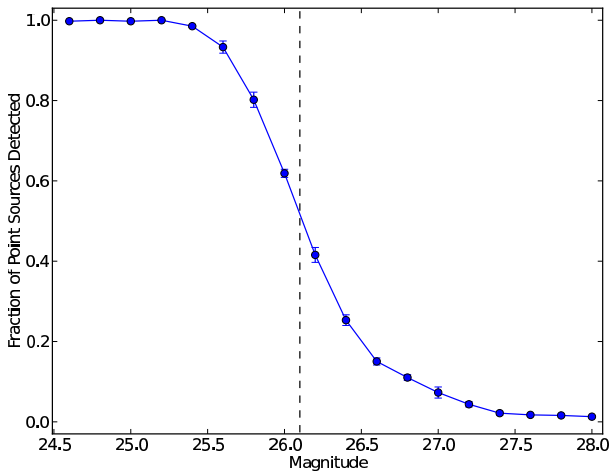}
\caption{Fraction of simulated point sources recovered by SExtractor,
  binned as a function of magnitude.  We are able to recover over 50
  per cent of sources in each magnitude bin, up to and including the
  $m_{K^{\prime}} =$~26.1 bin (represented by the dashed line).  We
  therefore treat this as our limiting magnitude.}
\label{fig:plimit}
\end{center}
\end{figure}

We generate a total of 1000 simulated images, randomizing both the
magnitudes and positions of the point sources, and then use SExtractor
\citep{ber96} to detect the added satellites.  We consider a point
source to be detected if SExtractor locates an object within 5 pixels
and 1.5 magnitudes of the actual position and magnitude of the
simulated object.  Compiling the results into magnitude bins, we find
that we are able to recover the majority (i.e. $>$~50~per cent) of the
simulated objects down to an apparent magnitude of K$^\prime=$~ 26.1,
which we take to be our detection limit (see Fig.  \ref{fig:plimit}).
Converting this value to the rest-frame V-band corresponds to an
object that has an absolute magnitude of $M_V = -$16.2, which is
fainter than the Magellanic clouds ($M_{V, \rm LMC} = -$18.5, $M_{V,
  \rm SMC} = -$17.1), but approximately three times brighter than the
Sagittarius dwarf satellite ($M_{V, \rm Sgr} = -$15; \citealt{tol08}).

\subsection{A second lensed source?}
\label{jackpot}

\begin{figure*}
\begin{center}
\centerline{
\includegraphics[width=5.8cm]{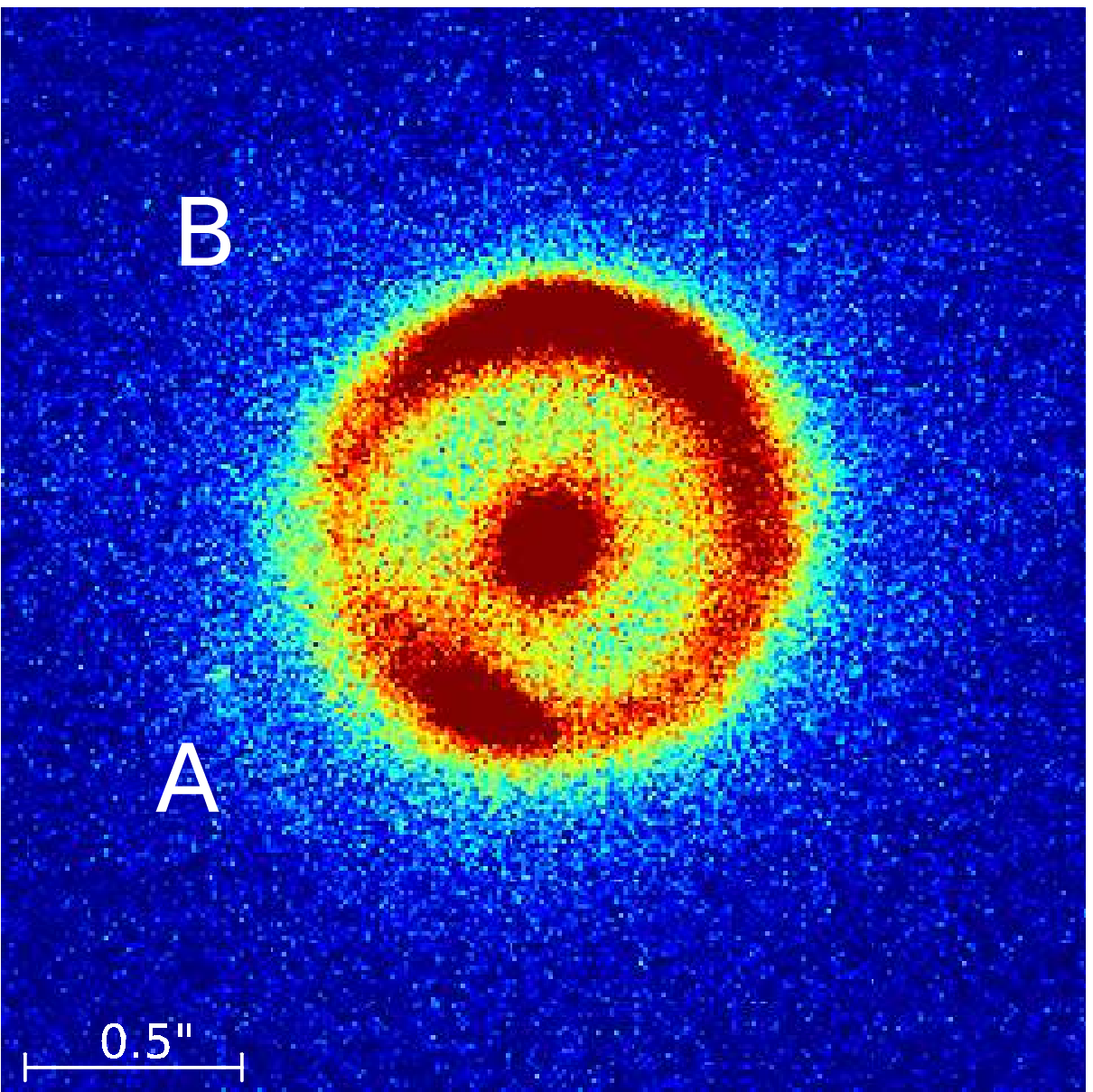}
\includegraphics[width=5.8cm]{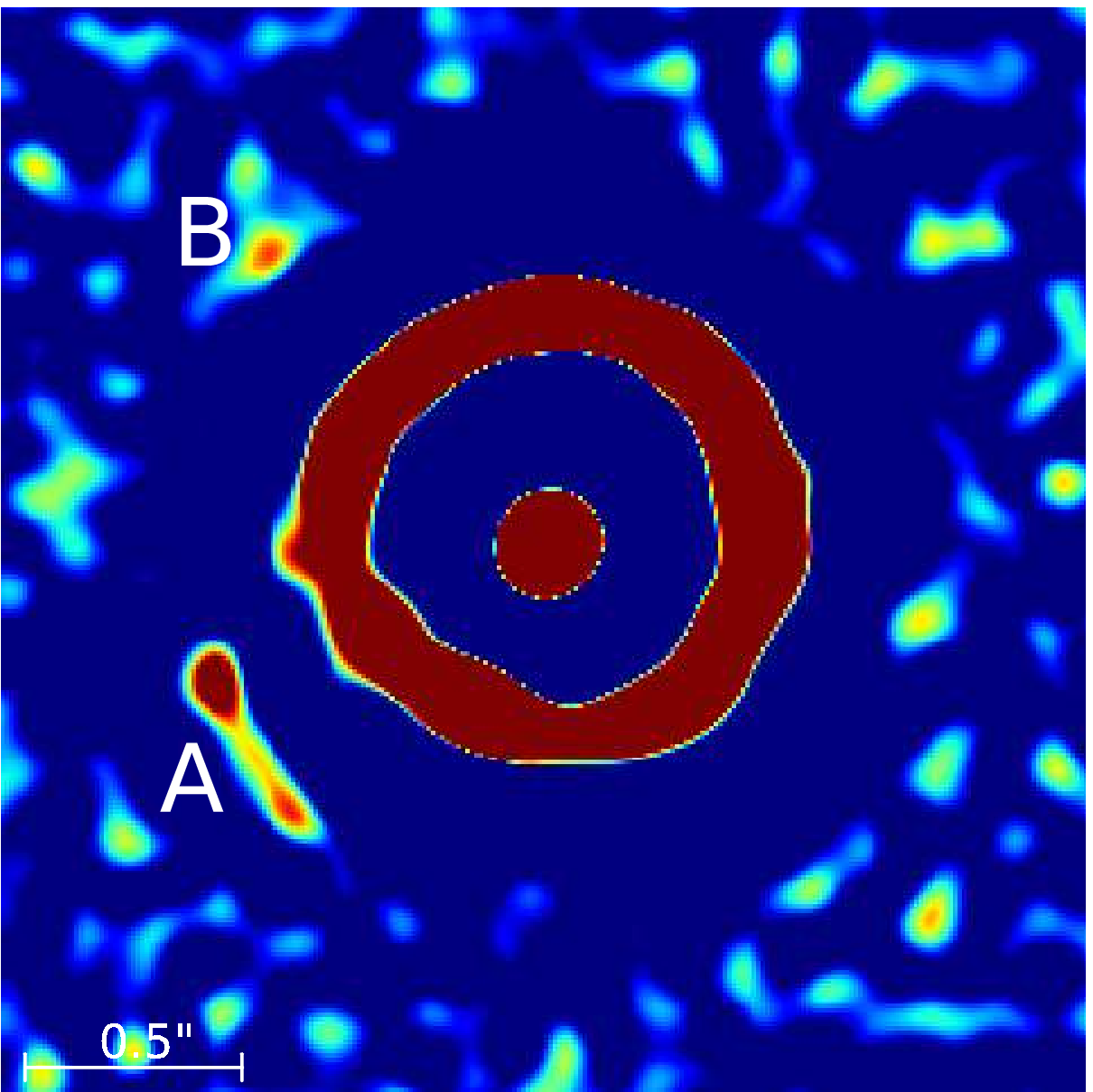}}
\end{center}
\caption{Left: K$^{\prime}$-band image of \lens\ with the two faint
  gravitational arc-like structures highlighted (A and B).  Right:
  same image after applying a Gaussian filter to smooth the image, and
  then a Laplacian filter to highlight the faint features.  Both
  potential gravitational arcs can clearly be seen.}
\label{fig:arcs}
\end{figure*}

There are two faint arclet structures that can be seen just outside
the Einstein ring on the eastern side of the K$^{\prime}$ image
(Fig. \ref{fig:imgs}; bottom middle panel).  To enhance the signal of
these arclets, we apply a Gaussian filter ($\sigma =$~6.5 pixels) to
smooth the data, followed by a Laplacian filter to increase the
contrast around bright flux peaks. Fig.  \ref{fig:arcs} shows a
high-contrast image of the original data and an image showing the
results of applying Gaussian and Laplacian filters. While the two arcs
are barely detectable above the noise in the unprocessed data, the
arcs are clearly visible in the filtered image.  As the arclets are so
faint in the original K$^{\prime}$-band data, we do not formally
consider them in our lens model.  However, taking into account the
enhanced image, it is possible -- given their positions and
orientations relative to the lensing galaxy -- that these arclets are
actually multiple images of a second background source.

There are three possible scenarios: the arclets are not being lensed;
the arclets are images of a strongly-lensed source that is associated
with the primary ring (e.g., an additional component of the galaxy
being lensed into the ring, a nearby galaxy, or more images of the
ring -- although this would lead to an extremely peculiar image
configuration that we do not expect to arise from well-behaved lensing
mass distributions); or the arclets are images of a second source
being strongly lensed but at a completely different redshift than the
Einstein ring galaxy. We can investigate these scenarios by using the
lens models that have been fitted to the Einstein ring. We first use
our best lens model to cast the emission from the Einstein ring and
the two arclets back to the source plane. As expected, the arclets do
not fall on the Einstein ring source. However, the two arclets
also are not mapped to a coincident location on the source plane,
indicating that if they are at the same redshift as the Einstein ring
galaxy they are not multiple images of the same source.

The possibility that the arclets correspond to images of a
strongly-lensed source at a different redshift is intriguing; such
double-plane lens systems provide very tight constraints to the
mass-density slope of the lens and can give useful constraints on
cosmological parameters (see \citealt{gav08} for details). To
investigate this possibility we must rescale the lens strength of our
mass model, where the scaling is directly proportional to the ratio of
angular diameter distances between the lens and the two sources. We
find that the two arclets can be mapped to approximately the same
location in the source plane if the lens strength is increased by a
factor of $\sim$~1.85. Such a large scaling factor requires an
unrealistically high redshift for the source, although two caveats
exist that could mitigate this concern: the galaxy being lensed into
the Einstein ring can contribute to the lensing, and the mass profile
could deviate from the central power law that was found from the
Einstein ring fit.

In any case, some estimate of the redshifts of the two potential
gravitational arcs will need to be obtained to determine their
nature. This will likely require extensive followup with deep,
high-resolution spectroscopy.  While this is difficult with current
telescopes (especially because arclet B is so faint), future
instruments such as the \textit{James Webb Space Telescope}
(\textit{JWST}), the Thirty Meter Telescope (TMT), or the
European-Extremely Large Telescope (E-ELT) should provide the
resolution and light-gathering power necessary to achieve this goal.

\section{Conclusions}
\label{conclusions}

We have presented a new mass model for the gravitational lens system
\lens, using a grid-based Bayesian reconstruction technique on high
resolution ground-based AO and space-based \emph{HST} data, as part of
the Strong-lensing at High Angular Resolution Program (SHARP).  We
find that the smooth component of the lensing galaxy's mass profile is
well-fitted by a (nearly isothermal) power-law distribution, while its
light profile (and that of the source galaxy) can be described by
S\'{e}rsic components.  A more in-depth analysis of the mass model,
characterizing the amount of substructure present in the system, is
presented in a companion paper to this work \citep{veg12}.  The model
is consistent over three independent NIR data sets -- due largely to
the constraints provided by a bright Einstein ring -- and agrees well
with previously reported results.  When compared to models derived
from traditional ground-based imaging, though, the high-resolution
models are significantly more precise.

The relative improvement in precision varies from data set to data
set.  Thus, by generating both AO- and {\it HST}-based models, we have
provided a quantitative comparison between instruments.  Overall, we
find that the uncertainties on the model parameters derived from the
AO data sets are smaller than those measured from the \emph{HST}, by
as much as an order of magnitude.  This suggests that AO data are
better at constraining lens models than equivalent \emph{HST} data (at
least in cases where the lens and source galaxies are red and have
reasonably smooth light profiles), and therefore, that AO observations
of lens systems can lead to a better description of those systems'
mass distributions.  However, confirmation of this result will require
a much larger data set and will be explored in a future paper (SHARP
II; Fassnacht et al., in prep).

\subsection*{ACKNOWLEDGMENTS}
DJL and CDF acknowledge support from NSF-AST-0909119. The Centre for
All-sky Astrophysics is an Australian Research Council Centre of
Excellence, funded by grant CE110001020.  SV is supported by a
Pappalardo Fellowship at the Massachusetts Institute of Technology and
is grateful to the Helena Klyuver female visitor program for funding
her stay at ASTRON, during which part of this work was carried
out. LVEK is supported (in part) through an NWO-VIDI program subsidy
(project number 639.042.505). All of the authors would like to extend
thanks to ASTRON for the warm hospitality (and financial support)
during the SHARP workshop, held in Dwingeloo, the Netherlands. This
research was supported in part by the National Science Foundation
under Grant No. NSF-PHY11-25915. The data presented herein were
obtained at the W. M. Keck Observatory, which is operated as a
scientific partnership among the California Institute of Technology,
the University of California and the National Aeronautics and Space
Administration. The Observatory was made possible by the generous
financial support of the W. M. Keck Foundation. The authors wish to
recognize and acknowledge the very significant cultural role and
reverence that the summit of Mauna Kea has always had within the
indigenous Hawaiian community. We are most fortunate to have the
opportunity to conduct observations from this mountain. The results
present herein were also based on observations collected with the
NASA/ESA {\it HST}, obtained at STScI, which is operated by AURA,
under NASA contract NAS5-26555.

\end{document}